\documentclass[twocolumn,showpacs,preprintnumbers,superscriptaddress,amsmath,amssymb]{revtex4}

\usepackage{graphicx}
\usepackage{dcolumn}
\usepackage{bm}
\usepackage{amssymb}

\newcommand{\thalf}{\tfrac{1}{2}}

\newcommand{\Tr}{\mbox{Tr}}

\begin{document}

\title{Continuity equation and local gauge invariance for the N$^{3}$LO
nuclear Energy Density Functionals}
\author{F. Raimondi}
 \email{francesco.raimondi75@gmail.com}
\affiliation{%
Department of Physics, P.O. Box 35 (YFL)
    FI-40014 University of Jyv\"askyl\"a, Finland.
}%

\author{B. G. Carlsson}%
\affiliation{Department of Physics, Lund University,
 P.O. Box 118
Lund
22100,
Sweden.}

\author{J. Dobaczewski}
\affiliation{%
Department of Physics, P.O. Box 35 (YFL)
    FI-40014 University of Jyv\"askyl\"a, Finland.
}%
\affiliation{Institute of Theoretical Physics, Faculty of Physics, University of Warsaw,
ul. Ho\.{z}a 69, PL-00-681 Warsaw, Poland.}

\author{J.~Toivanen}
\affiliation{%
Department of Physics, P.O. Box 35 (YFL)
    FI-40014 University of Jyv\"askyl\"a, Finland.
}%

\date{\today}

\begin{abstract}
\begin{description}
\item[Background:] The next-to-next-to-next-to-leading order (N$^{3}$LO) nuclear
energy density functional extends the standard Skyrme functional with
new terms depending on higher-order derivatives of densities,
introduced to gain better precision in the nuclear many-body
calculations.  A thorough study of the transformation properties of
the functional with respect to different symmetries is required, as a
step preliminary to the adjustment of the coupling constants.

\item[Purpose:] Determine to which extent the presence of
higher-order derivatives in the functional can be compatible with the
continuity equation. In particular, to study the relations between
the validity of the continuity equation and invariance of the
functional under gauge transformations.

\item[Methods:] Derive conditions for the validity of the continuity
equation in the framework of time-dependent density functional
theory. The conditions apply separately to the four
spin-isospin channels of the one-body density matrix.

\item[Results:] We obtained four sets of constraints on the coupling
constants of the N$^{3}$LO energy density functional that guarantee
the validity of the continuity equation in all spin-isospin channels.
In particular, for the scalar-isoscalar channel, the constraints are
the same as those resulting from imposing the standard U(1)
local-gauge-invariance conditions.

\item[Conclusions:] Validity of the continuity equation in the four
spin-isospin channels is equivalent to the local-gauge invariance of
the energy density functional. For vector and isovector channels,
such validity requires the invariance of the functional under local
rotations in the spin and isospin spaces.

\end{description}
\end{abstract}


\maketitle

\section{\label{sec:Intro}introduction}

In recent years, methods using energy density functionals (EDFs)
\cite{[Ben03]} to describe nuclear properties are being
developed in three complementary directions. First, the ideas of
effective theories~\cite{[Lep97],[Phi02]} are employed in determining
the EDFs from first
principles~\cite{[Fur08a],[Dru10],[Bog10],Stoitsov1}. These
developments are supplemented by a renewed
interest~\cite{Dobaczewski1,CarlssonPRL,Gebremariam1} in the
density-matrix expansion (DME) methods~\cite{Negele1,Negele2}, which
allow for treating exchange correlations in terms of (quasi)local
functionals. Second, the coupling constants of the well-known EDFs
undergo a thorough scrutiny, including an advanced work on the
readjustment of parameters~\cite{Erler1,[Kor10b]} and study of
inter-parameter correlations~\cite{[Rei10a]}. Finally, the standard
functionals are extended by adding new
terms~\cite{Carlsson1,[Zal10a],[Fan11],Raimondi1}, so as to gain
increased precision of description and predictability, in quest for
the spectroscopic-quality~\cite{[Zal08]} and universal~\cite{SCIDAC}
EDFs.

In the present work we study properties of EDFs~\cite{Carlsson1} and
pseudopotentials~\cite{Raimondi1} extended by adding terms that
depend on higher-order derivatives up to sixth,
next-to-next-to-next-to-leading order (N$^{3}$LO). Such extensions
lead to self-consistent mean-field Hamiltonians that are sixth-order
differential operators~\cite{[Car10d]}, that is, they depend on up to
sixth power of the momentum operator. This makes them unusual
objects, in the sense that standard second-order one-body
Hamiltonians contain only the Laplace operator in the kinetic-energy
term and possibly the angular-momentum operator in the spin-orbit
term. The main question we address here is whether the presence of
higher powers of momenta is compatible with the continuity equation
(CE).

The CE is a differential equation that describes a conservative
transport of some physical quantity~\cite{[Riley]}. In quantum
mechanics, it relates the time variation of the probability density
to the probability current \cite{[Mes62a]}. In our case, it appears
when the N$^{3}$LO EDFs or pseudopotentials are employed within a
time-dependent theory. For the standard Skyrme (NLO) functional,
the validity of the CE has been checked explicitly~\cite{Engel1}.
Our goal here is to derive constraints on the
coupling constant of the N$^{3}$LO EDF or parameters of the pseudopotential
that would guarantee the validity of the CE. Apart from linking the
CE to the local gauge symmetry~\cite{[Bla86]}, we also analyze the
CEs in vector and isovector channels and link them to the local
non-abelian gauge symmetries.

The paper is organized as follows. In Sec.~\ref{sec:the-CE1} we
present the standard quantal CE for a single particle and introduce
the vector CE. Then, in Sec.~\ref{sec:the-CE2} we discuss the CEs
within the time-dependent density functional theory and in
Sec.~\ref{sec:the-CE3} we specify the case to
the N$^3$LO quasilocal functional. The main body of
results obtained for the CEs in
the four spin-isospin channels is presented in Sec.~\ref{sec:Results}
and Appendices~\ref{app-scalar_isovector}--\ref{app-vector_isovector}.
Finally in Sec.~\ref{sec:Conclusions} we formulate the conclusions of
the present study.

\section{\label{sec:the-CE}Continuity equation in the EDF approach}

\subsection{\label{sec:the-CE1}Time evolution of a spin-{\boldmath$\thalf$} particle}

We begin by recalling the well-known \cite{[Mes62a]} CE for a single particle.
The time evolution of a non-relativistic spin-$\thalf$ particle moving
in a local potential is given by the Schr\"odinger equation,
\begin{eqnarray}
i\hbar \frac{\partial}{\partial t} \psi(\bm{r}\sigma,t)
 &=&  - \frac{\hbar^2}{2m}\Delta  \psi(\bm{r}\sigma,t)
  + V_0(\bm{r},t)\psi(\bm{r}\sigma,t)
\nonumber \\ \label{eq:1j}
 &&\hspace*{-1.5cm}+ \sum_{\mu=x,y,z}V_{1\mu}(\bm{r},t)
\sum_{\sigma'}\langle\sigma|\sigma_\mu|\sigma'\rangle\psi(\bm{r}\sigma',t),
\end{eqnarray}
where $ V_0(\bm{r},t)$ and $V_{1\mu}(\bm{r},t)$ are scalar and vector
real time-dependent potentials, respectively, and
$\langle\sigma|\sigma_\mu|\sigma'\rangle$ are the standard Pauli
matrices. By multiplying Eq.~(\ref{eq:1j}) with $\psi^*(\bm{r}\sigma,t)$, summing up over
$\sigma$, and taking the imaginary part, we obtain the standard CE for the
probability
density $\rho(\bm{r},t)$ in terms of the current $\bm{j}(\bm{r},t)$,
\begin{equation}
\frac{\partial}{\partial t} \rho(\bm{r},t) = -\frac{\hbar}{m}\bm{\nabla}\cdot\bm{j}(\bm{r},t)
\label{eq:2j},
\end{equation}
where
\begin{subequations}
\begin{eqnarray}
\rho(\bm{r},t)   &=& \sum_{\sigma} |\psi(\bm{r}\sigma,t)|^2 , \\
\bm{j}(\bm{r},t) &=& \sum_{\sigma}
                   \mbox{Im}\Big(\psi^*(\bm{r}\sigma,t)\bm{\nabla}\psi  (\bm{r}\sigma,t)\Big) .
\end{eqnarray}
\end{subequations}
We see that the hermiticity of the local potential guarantees that the
potential energy does not contribute to the CE of Eq.~(\ref{eq:2j}).

Similarly, by multiplying Eq.~(\ref{eq:1j}) with
$\psi^*(\bm{r}\sigma'',t)\langle\sigma''|\sigma_\nu|\sigma\rangle$, summing up over
$\sigma''$ and $\sigma$, and taking the imaginary part, we obtain the CE for the
spin density $s_\nu(\bm{r},t)$ in terms of the spin current $\bm{J_\nu}(\bm{r},t)$,
\begin{eqnarray}\!\! \!\!
\frac{\partial}{\partial t} s_\nu(\bm{r},t) &\!\! =\!\! & - \frac{\hbar}{m}\bm{\nabla}\cdot\bm{J}_\nu(\bm{r},t)
 + \frac{1}{\hbar}\Big(\!\bm{V}_1(\bm{r},t)\!\!\times\!\!\bm{s}(\bm{r},t)\!\Big)_\nu ,
\label{eq:3j}
\end{eqnarray}
where
\begin{subequations}
\begin{eqnarray}
\!\!\!\!
s_\nu(\bm{r},t)   &\!\!=\!\!& \sum_{\sigma'\sigma}
  \psi^*(\bm{r}\sigma',t)\langle\sigma'|\sigma_\nu|\sigma\rangle\psi(\bm{r}\sigma,t) , \\
\!\!\!\!
\bm{J_\nu}(\bm{r},t) &\!\!=\!\!& \sum_{\sigma'\sigma}
                   \mbox{Im}\Big(\psi^*(\bm{r}\sigma',t)\langle\sigma'|\sigma_\nu|\sigma\rangle\bm{\nabla}\psi  (\bm{r}\sigma,t)\Big) .
\end{eqnarray}
\end{subequations}
We see that the spin CE does depend on the vector potential,
and the second term in Eq.~(\ref{eq:3j}) is responsible, e.g., for
the spin precession in magnetic field.

It is interesting to note that when potential $\bm{V}_1(\bm{r},t)$ is
parallel to the spin density $\bm{s}(\bm{r},t)$ (non-linear
Schr\"odinger equation), {\em all} components of the spin density
fulfill the CEs. In fact, this is exactly the case for the TDHF
equation induced by a zero-range two-body interaction, see below.
Another interesting case corresponds to the vector potential aligned
along a fixed direction in space, say, along the $z$ axis, that is
$\bm{V}_1(\bm{r},t)={V}_1(\bm{r},t)\bm{e}_z$. In this case, the time
evolutions of the spin-up and spin-down components decouple from one
another, that is, $\bm{s}(\bm{r},0)=s(\bm{r},0)\bm{e}_z$ implies
$\bm{s}(\bm{r},t)=s(\bm{r},t)\bm{e}_z$, and the spin-up and spin-down
components individually obey the corresponding CEs.

We also note here that for a nonlocal potential-energy term,
\begin{equation}
(\hat{V}\psi)(\bm{r}\sigma,t) = \int {\rm d}^3\bm{r}'\sum_{\sigma'}
V(\bm{r}\sigma,\bm{r}'\sigma',t)\psi(\bm{r}'\sigma',t)
\label{eq:4j},
\end{equation}
the time evolution does not, in general, lead to a CE.

\subsection{\label{sec:the-CE2}Time-dependent density functional theory}

In the framework of the time-dependent
Hartree-Fock (TDHF) approximation or time-dependent
density functional theory (TDDFT), the so-called
memory effects are often neglected and it is assumed that the potential at
time $t$ is just the static potential evaluated at the
instantaneous density~\cite{[Vignale]}. For these two time-dependent
approaches, the starting point is the equation of motion
for the one-body density matrix $\rho_{\alpha\beta}$ \cite{[RS80],[Bla86]},
\begin{equation}
i\hbar \frac{{\rm d}}{{\rm d} t}\rho = [h,\rho]
\label{eq:5j},
\end{equation}
where the mean-field Hamiltonian $h_{\alpha\beta}$ is defined as the derivative
of the total energy $E \{\rho\}$ with respect to the density matrix,
\begin{equation}
h_{\alpha\beta} = \frac{\partial E \{\rho\}}{\partial \rho_{\beta\alpha}}
\label{eq:6j}.
\end{equation}

In the present study we are concerned with the Kohn-Sham approach \cite{[Koh65a]},
whereby the total energy is the sum of the kinetic and potential-energy terms,
\begin{equation}
E \{\rho\} = E_k \{\rho\} + E_p \{\rho\}
\label{eq:7j},
\end{equation}
where
\begin{equation}
E_k \{\rho\} = \frac{\hbar^2}{2m}\int {\rm d}^3\bm{r} \tau_0^0({\bm r},t)
\label{eq:8j}
\end{equation}
and $\tau_0^0({\bm r},t)=\Big(\sum_{\sigma\tau}{\bm\nabla}\cdot{\bm\nabla'}
\rho({\bm r}\sigma\tau,{\bm r}'\sigma\tau,t)\Big)|_{{\bm r}={\bm r}'}$
is the scalar-isoscalar kinetic density, see, e.g.,\ Ref.~\cite{[Per04]}
for definitions. The nonlocal density,
can be defined in terms of either the fixed-basis orbitals, $\psi_\alpha({\bm r}\sigma\tau)$,
\begin{eqnarray}
\!\!\!\!\!\!\!\!\rho({\bm r}\sigma\tau,{\bm r}'\sigma'\tau',t)
&\!\!=\!\!&\sum_{\beta\alpha}\psi_\beta({\bm r}\sigma\tau)\rho_{\beta\alpha}(t)
\psi^*_\alpha({\bm r}'\sigma'\tau')
\label{eq:15ja},
\end{eqnarray}
or instantaneous Kohn-Sham orbitals, $\phi_i({\bm r}\sigma\tau,t)$,
\begin{eqnarray}
\!\!\!\!\!\!\!\!\rho({\bm r}\sigma\tau,{\bm r}'\sigma'\tau',t)
&\!\!=\!\!&\sum_{i=1}^A\phi_i({\bm r}\sigma\tau,t)
\phi^*_i({\bm r}'\sigma'\tau',t)
\label{eq:15jb}.
\end{eqnarray}
The mean-field Hamiltonian
is the sum of kinetic and potential-energy terms,
$h_{\alpha\beta}=T_{\alpha\beta}+\Gamma_{\alpha\beta}$, where
\begin{equation}
T_{\alpha\beta} = \int {\rm d}^3\bm{r}\sum_{\sigma\tau}\psi^*_\alpha({\bm r}\sigma\tau)
\frac{- \hbar^2}{2m}\Delta\psi_\beta({\bm r}\sigma\tau)
\label{eq:9j}
\end{equation}
and
\begin{equation}
\Gamma_{\alpha\beta} = \frac{\partial E_p \{\rho\}}{\partial \rho_{\beta\alpha}}
\label{eq:10j}.
\end{equation}

Let us now assume that the potential energy is invariant with respect
to a unitary transformation of the density matrix \cite{[RS80],[Bla86]},
$U=\exp(i\eta G)$, that is,
for all parameters $\eta$ we have,
\begin{equation}
E_p \{\rho\} = E_p \{U\rho U^+\}
\label{eq:11j},
\end{equation}
where $G_{\alpha\beta}$ is the hermitian matrix of a one-body symmetry generator.
Then, the first-order expansion in $\eta$,
\begin{equation}
E_p \{U\rho U^+\} \simeq  E_p \{\rho\}
 +\eta \sum_{\beta\alpha}\left[\frac{\partial E_p \{\rho\}}{\partial \rho_{\beta\alpha}}
\frac{\partial (U\rho U^+)_{\beta\alpha}}{\partial \eta}\right]_{\eta=0}
\label{eq:12j},
\end{equation}
gives a condition for the energy to be invariant with respect to this unitary transformation, that is
\begin{equation}
\Tr \Gamma[G,\rho] \equiv \Tr G[\Gamma,\rho] = 0
\label{eq:13j},
\end{equation}
which allows us to derive the equation of motion for the average value
of $\langle G\rangle=\Tr G\rho$. Indeed, from the TDDFT equation (\ref{eq:5j})
we then have:
\begin{equation}
i\hbar \frac{{\rm d}}{{\rm d} t}\langle G\rangle
= i\hbar \Tr G \frac{{\rm d}}{{\rm d} t}\rho = \Tr G[h,\rho] = \Tr G[T,\rho]
\label{eq:14j},
\end{equation}
that is, the time evolution of $\langle G\rangle$ is governed solely
by the kinetic term of the mean-field Hamiltonian.

\subsubsection{\label{sec:the-CE2-00}Continuity equation for the scalar-isoscalar density}

The CE now results from specifying $\eta G$ to the
local gauge transformation~\cite{[Bla86],Dobaczewski2} that is defined as
\begin{equation}
\psi'_\alpha({\bm r}\sigma\tau) \equiv (U\psi_\alpha)({\bm r}\sigma\tau)
=e^{i\gamma({\bm r})}\psi_\alpha({\bm r}\sigma\tau)
\label{eq:16j}.
\end{equation}
Then, Eq.~(\ref{eq:15ja}) gives:
\begin{equation}
\rho'({\bm r}\sigma\tau,{\bm r}'\sigma'\tau',t)
= e^{i\left(\gamma({\bm r})-\gamma({\bm r}')\right)}
\rho({\bm r}\sigma\tau,{\bm r}'\sigma'\tau',t)
\label{eq:17j}.
\end{equation}
Matrix elements of the local-gauge angle $\gamma({\bm r})$ are
given by local integrals,
\begin{equation}
\gamma_{\alpha\beta} = \int {\rm d}^3\bm{r}\sum_{\sigma\tau}\psi^*_\alpha({\bm r}\sigma\tau)
\gamma({\bm r})\psi_\beta({\bm r}\sigma\tau)
\label{eq:18j};
\end{equation}
therefore, from Eq.~(\ref{eq:15ja}) again, the average value of the
gauge angle, $\langle\gamma\rangle=\Tr\gamma\rho$, depends on the scalar-isoscalar local density
$\rho_0^0({\bm r},t)=\sum_{\sigma\tau}\rho({\bm r}\sigma\tau,{\bm r}\sigma\tau,t)$,
that is,
\begin{equation}
\langle\gamma\rangle = \int {\rm d}^3\bm{r}
\gamma({\bm r})\rho_0^0({\bm r},t)
\label{eq:19j}.
\end{equation}

Now, the assumed local-gauge invariance of the potential energy
implies the equation of motion for the average value $\langle\gamma\rangle$,
which from Eq.~(\ref{eq:14j}) reads
\begin{equation}
\frac{{\rm d}}{{\rm d} t}\langle\gamma\rangle
 = -\frac{\hbar}{m}\int {\rm d}^3\bm{r}\gamma({\bm r})
\bm{\nabla}\cdot\bm{j}_0^0(\bm{r},t)
\label{eq:20j},
\end{equation}
where the standard scalar-isoscalar current is defined as \cite{[Per04]}
$\bm{j}_0^0({\bm r},t)=\sum_{\sigma\tau}\frac{1}{2i}
\left[({\bm\nabla}-{\bm\nabla}')
\rho({\bm r}\sigma\tau,{\bm r}'\sigma\tau,t)\right]_{{\bm r}={\bm r}'}$.

We note here~\cite{[Bla86],Dobaczewski2}, that the gauge invariance
that corresponds to a specific dependence of the gauge angle on
position, $\gamma({\bm r})={\bm P}_0\cdot{\bm r}$, represents the
Galilean invariance of the potential energy for the system boosted to
momentum ${\bm P}_0$. Then, equation of motion (\ref{eq:20j}) simply
represents the classical equation for the center-of-mass velocity,
\begin{equation}
\frac{{\rm d}}{{\rm d} t}\frac{\langle{\bm r}\rangle}{A}
 \equiv\frac{{\rm d}}{{\rm d} t}{\bm R}_{\text{CM}}
 = \frac{\langle{\bm P}\rangle}{mA}
 \equiv \frac{\langle -i\hbar{\bm\nabla}\rangle}{mA}
\label{eq:22j}.
\end{equation}

In the general case, that is, when the potential energy is gauge-invariant
and the gauge angle $\gamma({\bm r})$ is an {\em arbitrary} function of ${\bm r}$,
Eq.~(\ref{eq:20j}) gives the CE that reads
\begin{equation}
\frac{{\rm d}}{{\rm d} t} \rho_0^0({\bm r},t)
 = -\frac{\hbar}{m} \bm{\nabla}\cdot\bm{j}_0^0(\bm{r},t)
\label{eq:21j}.
\end{equation}
Thus for a gauge-invariant potential energy
density, the TDHF or TDDFT equation of motion implies the CE, that is,
the gauge invariance is a sufficient condition for the validity of the CE.
By proceeding in the opposite direction, we can prove that it is also a necessary
condition. Indeed, the CE of Eq.~(\ref{eq:21j}) implies the first-order
condition (\ref{eq:13j}), and then the full gauge invariance stems from
the fact that the gauge transformations form local U(1) groups.

\subsubsection{\label{sec:the-CE2-vt}Continuity equation for densities in spin-isospin channels}

We can now repeat derivations presented in
Eqs.~(\ref{eq:16j})-(\ref{eq:20j}) by considering the spin-isospin
local-gauge groups, and derive CEs in other spin-isospin channels.
To this end, we first express the nuclear one-body density
matrix (\ref{eq:15ja})--(\ref{eq:15jb})
as a linear combination of nonlocal
spin-isospin densities $\rho_v^t(\bm{r},\bm{r}')$~\cite{CarlssonPRL},
\begin{eqnarray}
\nonumber && \hspace*{-1cm}\rho(\bm{r}\sigma \tau, \bm{r}'\sigma' \tau') = \\
&&  \frac{1}{4} \sum_{\substack{v=0,1 \\ t=0,1}}\left(\sqrt{3}\right)^{v+t}
\left[\sigma_v^{\sigma\sigma'} \left[\tau_{\tau\tau'}^t  \rho_v^t(\bm{r},\bm{r}') \right]^0\right]_0
\label{eq:1bis},
\end{eqnarray}
where the sums run over the spin ($v=0,1$) and isospin ($t=0,1$)
indices denoted by subscripts and superscripts, respectively, coupled
to total scalar and isoscalar. Here and below we use the coupling of
spherical tensors both for angular momentum and isospin tensors;
therefore, in Eq.~(\ref{eq:1bis}) the factor of
$\left(\sqrt{3}\right)^{v+t}$ was included so as to cancel the
corresponding values of the Clebsch-Gordan coefficients, and to
maintain the standard normalization of the spin-isospin densities.
The spin-isospin densities can be conversely expressed as the following
traces of the density matrix,
\begin{eqnarray}
\rho_v^t(\bm{r},\bm{r}') &=& \sum_{\sigma \tau,\sigma' \tau'}
\sigma_v^{\sigma'\sigma}\tau_{\tau'\tau}^t
\rho(\bm{r}\sigma \tau, \bm{r}'\sigma' \tau')
\label{eq:1bis2}.
\end{eqnarray}

The CEs for densities in the
scalar-isoscalar ($v=0$, $t=0$), scalar-isovector ($v=0$, $t=1$),
vector-isoscalar ($v=1$, $t=0$), and vector-isovector ($v=1$, $t=1$) channels,
\begin{equation}
\frac{{\rm d}}{{\rm d} t} \rho^t_v({\bm r})
 = -\frac{\hbar}{m} \bm{\nabla}\cdot\bm{J}^t_v(\bm{r})
\label{eq:702},
\end{equation}

\vspace*{2mm}\noindent
where
$\bm{J}^t_v(\bm{r})=\frac{1}{2i}\left(\bm{\nabla}-\bm{\nabla}'\right)
\rho^t_v(\bm{r},\bm{r}')|_{\bm{r}'=\bm{r}}$ and
$\rho^t_v(\bm{r})=\rho^t_v(\bm{r},\bm{r})$, are now equivalent to the
local gauge invariances, respectively, with respect to the four local
spin-isospin groups:
\begin{eqnarray}
\label{eq:731a}
U^t_v(\bm{r}) &=& \exp\left(i\left[\left[\gamma^t_v(\bm{r})\sigma_v\right]_0\tau^t\right]^0\right) .
\end{eqnarray}
Of course, the standard CE derived in Sec.~\ref{sec:the-CE2-00}
corresponds to $\gamma(\bm{r})\equiv\gamma^0_0(\bm{r})$. Note that the
four gauge groups are different: $U^0_0(\bm{r})$ gives the standard
abelian gauge group U(1), $U^0_1(\bm{r})$ and $U^1_0(\bm{r})$ form
the non-abelian gauge groups SU(2), whereas $U^1_1(\bm{r})$
corresponds to the non-abelian gauge group SU(2)$\times$SU(2).

\subsection{\label{sec:the-CE3}The N{\boldmath$^{3}$}LO quasilocal functional}

We are now in a position to discuss the CE for the N$^{3}$LO
quasilocal functional introduced by Carlsson
{\it et al.}~\cite{Carlsson1}. By imposing on the functional the
gauge-invariance conditions, we can then confirm and explicitly
rederive the results of Sec.~\ref{sec:the-CE2}. The explicit
derivation will also allow us to discuss the CEs for densities in
other spin-isospin channels
analyzed in Sec.~\ref{sec:the-CE2-vt}.

Below we consider the EDF given in terms of a local integral
of the energy density $\mathcal{H}_E(\bm{r})$,
\begin{equation}
E\{\rho\} = \int {\rm d}^3\bm{r} \ \mathcal{H}_E(\bm{r})
\label{eq:0},
\end{equation}
which is represented as a sum of the kinetic and potential energies
conforming to Eq.~(\ref{eq:7j}),
\begin{equation}
 \mathcal{H}_E(\bm{r})= \frac{\hbar^{2}}{2 m} \tau_0^0(\bm{r}) +
\sum_{t=0,1} \mathcal{H}^t(\bm{r})
\label{eq:1}.
\end{equation}
To lighten the notation and avoid confusion with
the isospin index $t=0,1$, in this section we do not explicitly show the
time argument of densities, which within the TDDFT all depend on time.

The quasilocal N$^{3}$LO EDF was constructed~\cite{Carlsson1} by
building the $t=0$ and $t=1$ potential-energy densities $\mathcal{H}^t(\bm{r})$
from isoscalar and isovector densities, respectively,
and their derivatives up to sixth order. For clarity, we give here
a brief summary of definitions and notations used in this
construction.

The local higher-order primary densities are defined by the coupling
of relative-momentum tensors $K_{n L}$~\cite{Carlsson1} with nonlocal
densities (\ref{eq:1bis2}) to total angular momentum ${J}$, that is,
 \begin{equation}
 \rho_{n L v J}^t(\bm{r})=  \left\{ \left[ K_{n L} \rho_{v}^t(\bm{r}, \bm{r}')  \right]_{J} \right\}_{\bm{r}'= \bm{r}}
 \label{eq:3}.
 \end{equation}
Then, a general term of the N$^{3}$LO functional can be written, in
the language of the spherical tensors, as
\begin{eqnarray}
T_{m I, n L v J}^{n' L' v' J',t} (\bm{r}) \!\!=\!\!
\left[\left[ \rho_{n' L' v' J'}^t (\bm{r})
\left[D_{m I} \rho_ {n L v J}^t(\bm{r}) \right]_{J'} \right]^{0} \right]_{0}\!\!
\label{eq:2},
\end{eqnarray}
where the local secondary densities,
$\left[D_{m I} \rho_ {n L v J}^t(\bm{r})\right]_{J'}$, are obtained
by acting with derivatives $D_{m I}$ on primary densities and coupling them
to total angular momentum ${J'}$. Each term (\ref{eq:2}) is multiplied
by the corresponding coupling constant $C_{m I, n L v J}^{n' L' v' J',t}$ that is
denoted by the same set of indices as those in the term itself.

We note here that the definition of the isovector terms depends on
whether one uses Cartesian or spherical
representation of tensors in isospace. On the one hand, the use of the standard
Cartesian representation, see, e.g., Refs.~\cite{[Per04],Carlsson1},
implies that the isovector terms depend on products of differences of
neutron and proton densities. On the other hand, the use of the
spherical representation, which was assumed in Ref.~\cite{Raimondi1}
and is also used in the present study, involves the coupling of two
isovectors to a scalar, whereby there appears a Clebsch-Gordan
coefficient of $\left(\sqrt{3}\right)^{-1}$. Therefore, for the
isospace spherical representation, the isovector coupling constants
are by the factor of $\left(\sqrt{3}\right)$ {\em larger} than those
for the Cartesian representation.

In the remaining part of this section, we employ the compact notation
introduced in Ref.~\cite{[Car10d]}, whereby the grouped indices, such
as the Greek indices $\alpha=\left\{
n_{\alpha}L_{\alpha}v_{\alpha}J_{\alpha}\right\}$ and the Roman
indices $a=\left\{m_aI_a\right\}$, denote all the quantum numbers of
the local densities $\rho_{\alpha}(\bm{r})$ and derivative operators
$D_{a}$, respectively. In this notation, the N$^{3}$LO
potential-energy density of Eq.~(\ref{eq:1}) reads
 \begin{equation}
\label{eq:hfeq2}
{\cal H}^t(\bm{r})=\sum_{a\alpha\beta}C_{a,\alpha}^{\beta,t}T_{a,\alpha}^{\beta,t}(\bm{r}).
 \end{equation}

Our following discussion of the CE is mainly focused on the
one-body potential-energy term, defined in Eq.~(\ref{eq:10j}) as
the variation of the potential energy with respect to the density
matrix. For the N$^{3}$LO functional, this term was derived
in Ref.~\cite{[Car10d]}, where it was shown that in space coordinates
it has the form of a one-body pseudopotential,
\begin{equation}
\hat{\Gamma}^{\sigma\sigma'}_{\tau\tau'}(\bm{r})=
\sum_{\gamma,t}\left[\left[U_{\gamma}^{t}(\bm{r})
\left[D_{n_{\gamma}L_{\gamma}}\sigma^{\sigma\sigma'}_{v_{\gamma}}
\right]_{J_{\gamma}}\right]_{0}\tau^{t}_{\tau\tau'}\right]^{0}
\label{eq:7}.
\end{equation}
An equivalent form of the one-body pseudopotential, which can be obtained
by recoupling spherical tensors within a scalar, and which
separates out the spin Pauli matrices, reads
\begin{equation}
\hat{\Gamma}^{\sigma\sigma'}_{\tau\tau'}(\bm{r})=
\sum_{\gamma,t}\left[\left[\left[U_{\gamma}^{t}(\bm{r})
D_{n_{\gamma}L_{\gamma}}\right]_{v_{\gamma}}\sigma^{\sigma\sigma'}_{v_{\gamma}}\right]_{0}
\tau^{t}_{\tau\tau'}\right]^{0}
\label{eq:7s}.
\end{equation}

In turn, potentials $U_{\gamma}^{t}(\bm{r})$ were derived as linear combinations
of the secondary densities,
\begin{equation}
U_{\gamma}^{t}(\bm{r})=\underset{a\alpha\beta;d\delta}{\sum}
C_{a,\alpha}^{\beta,t}\chi_{a,\alpha;\gamma}^{\beta;d\delta}
\left[D_{d}\rho^{t}_{\delta}(\bm{r})\right]_{J_{\gamma}}
\label{eq:7-1},
\end{equation}
where $\chi_{a,\alpha;\gamma}^{\beta;d\delta}$ are numerical
coefficients. We call the one-body operator
$\hat{\Gamma}^{\sigma\sigma'}_{\tau\tau'}(\bm{r})$ pseudopotential,
because it is defined in terms of potentials $U_{\gamma}^{t}(\bm{r})$
{\em and} differential operators $D_{n_{\gamma}L_{\gamma}}$ acting on
single-particle wave functions. We note here, that in
Eqs.~(\ref{eq:7}) and (\ref{eq:7s}) potentials always appear to the left of all
derivatives; nonetheless, the one-body pseudopotential is a
hermitian operator, which is guaranteed by specific conditions
obeyed by potentials $U_{\gamma}^{t}(\bm{r})$, which were derived in
Ref.~\cite{[Car10d]}.

For the one-body pseudopotential (\ref{eq:7}),
the Schr\"odinger equation that gives the time evolution of
single-particle Kohn-Sham wave functions in space coordinates reads,
\begin{eqnarray}
i\hbar\frac{\partial}{\partial t}\phi_{i}(\bm{r}\sigma\tau,t)
&=& -\frac{\hbar^{2}}{2m}\Delta\phi_{i}(\bm{r}\sigma\tau,t)
\nonumber \\
&+&\sum_{\sigma'\tau'}\hat{\Gamma}^{\sigma\sigma'}_{\tau\tau'}(\bm{r})
\phi_{i}(\bm{r}\sigma'\tau',t)
\label{eq:1c}.
\end{eqnarray}
By multiplying the Schr\"odinger equation with the complex-conjugated wave function,
$\phi^*_{i}(\bm{r}'\sigma'\tau',t)$ and summing over the single-particle index $i$
we obtain the time-evolution equation
of the density matrix (\ref{eq:15jb}), that is,
\begin{eqnarray}
&&\hspace{-1cm}i \hbar \frac{\partial}{\partial t}\rho(\bm{r} \sigma \tau, \bm{r}' \sigma' \tau' ,t) = \nonumber \\
&&-\frac{\hbar^2}{2m}\left(\Delta-\Delta'\right)
\rho(\bm{r} \sigma \tau, \bm{r}' \sigma' \tau' ,t) \nonumber \\
&& + \sum_{\sigma''\tau''}\left(
\hat{\Gamma}^{\sigma\sigma''}_{\tau\tau''}(\bm{r})
\rho(\bm{r} \sigma''\tau'', \bm{r}' \sigma' \tau' ,t)  \right. \nonumber \\
&&\hspace{0.9cm} - \left. \hat{\Gamma}^{\sigma'\sigma''*}_{\tau'\tau''}(\bm{r}')
\rho(\bm{r} \sigma \tau, \bm{r}' \sigma'' \tau'' ,t)\right)
\label{eq:7c}.
\end{eqnarray}

Before we proceed, we must first consider the complex-conjugated
pseudopotential
$\hat{\Gamma}^{\sigma'\sigma''*}_{\tau'\tau''}(\bm{r}')$. To this
end, we use the property of the Biedenharn-Rose phase convention employed in
Refs.~\cite{Carlsson1,[Car10d]}, by which all scalars are always real.
Note that for the spherical representation of Pauli matrices,
the Biedenharn-Rose phase convention implies the transposition
of spin indices, that is,
\begin{equation}
\left(\sigma^{\sigma\sigma'}_{v\mu}\right)^*
= (-1)^{v-\mu} \sigma^{\sigma'\sigma}_{v,-\mu}  ,
\label{eq:246}
\end{equation}
where $\mu=0$ for $v=0$ and $\mu=-1,0,1$ for $v=1$ denote tensor components
of scalar and vector Pauli matrices, respectively.

Finally, in Eqs.~(\ref{eq:7s}) and (\ref{eq:7-1}), the complex
conjugation only affects coefficients
$\chi_{a,\alpha;\gamma}^{*\beta;d\delta}=
(-1)^{n_\gamma+m_a+m_d}\chi_{a,\alpha;\gamma}^{\beta;d\delta}$~\cite{[Car10d]},
which gives,
\begin{equation}
\hat{\Gamma}^{\sigma'\sigma''*}_{\tau'\tau''}(\bm{r}') =
\hat{\Gamma}^{'\sigma''\sigma'}_{\tau''\tau'}(\bm{r}')
\label{eq:701},
\end{equation}
for
\begin{equation}
\hat{\Gamma}^{'\sigma\sigma'}_{\tau\tau'}(\bm{r}')=
\sum_{\gamma,t}\left[\left[\left[U_{\gamma}^{'t}(\bm{r}')
D'_{n_{\gamma}L_{\gamma}}\right]_{v_{\gamma}}\sigma^{\sigma\sigma'}_{v_{\gamma}}\right]_{0}
\tau^{t}_{\tau\tau'}\right]^{0}
\label{eq:700s}
\end{equation}
and
\begin{equation}
U_{\gamma}^{'t}(\bm{r}')=\underset{a\alpha\beta;d\delta}{\sum}
(-1)^{n_\gamma+m_a+m_d}
C_{a,\alpha}^{\beta,t}\chi_{a,\alpha;\gamma}^{\beta;d\delta}
\left[D'_{d}\rho^{t}_{\delta}(\bm{r}')\right]_{J_{\gamma}}
\label{eq:700-1}.
\end{equation}
It means that in all further derivations we must use the second set of
potentials $U_{\gamma}^{'t}(\bm{r}')$ with signs of terms
modified according to the phase $(-1)^{n_\gamma+m_a+m_d}$.
It is now obvious that the CEs will hold independently of the
spin-isospin coordinates if, and only if, the pseudopotentials
fulfill the condition
\begin{eqnarray}
&&\underset{\sigma''\tau''}{\sum}\left(\Gamma_{\tau\tau''}^{\sigma\sigma''}(\mathbf{r})\rho(\mathbf{r}\sigma''\tau'',\mathbf{r'}\sigma'\tau') \right.
\nonumber \\ && ~~~~~~ \left.
-\Gamma_{\tau''\tau'}^{'\sigma''\sigma'}(\mathbf{r}')\rho(\mathbf{r}\sigma\tau,\mathbf{r'}\sigma''\tau'')\right)_{\bm{r}'=\bm{r}}=0
\label{eq:8.2}.
\end{eqnarray}

We are now in a position to separate the four
spin-isospin channels in Eq.~(\ref{eq:7c}). We do so by multiplying both sides of the
equation with $\sigma_v^{\sigma'\sigma}\tau_{\tau'\tau}^t$ and
summing over $\sigma \tau,\sigma' \tau'$. From Eq.~(\ref{eq:1bis2})
it is then obvious that, in close analogy to Sec.~\ref{sec:the-CE1},
after setting $\bm{r}'=\bm{r}$, we obtain the CEs (\ref{eq:702})
in the four spin-isospin channels,
provided terms coming from one-body pseudopotentials do
not contribute, as in Eq.~(\ref{eq:8.2}). When evaluating this
condition for the four spin-isospin channels, we use the
expression for the trace of three Paul matrices in spherical
representation, which reads \cite{[Var88]},
\begin{equation}
{\rm Tr}\Big[\sigma_{v\mu}\sigma_{v'\mu'}\sigma_{v''\mu''}\Big]
=A(v+v'+v'')(-1)^{v-\mu}C^{v,-\mu}_{v'\mu'v''\mu''}  ,
\label{eq:235a}
\end{equation}
where we introduced $A(v+v'+v'')$ as a shorthand symbol for numerical coefficients coming from the computation
of the trace. In the calculation, we only need values of $A(0)=2$,  $A(2)=2\sqrt{3}$,  $A(3)=2\sqrt{2}\,i$.
After a trivial but lengthy calculation, we obtain the final result:
\begin{widetext}
\begin{eqnarray}
&& \hspace*{-1cm}\sum_{\substack{v't' \\ \gamma,t''J}}
A(v+v'+v_{\gamma})A(t+t'+t'')
\frac{(-1)^{v'+v+J_{\gamma}+L_{\gamma}}}{4\sqrt{3}^{t''}}
\sqrt{(2J+1)}
\left\{\begin{array}{ccc} J_{\gamma} & L_{\gamma} & v_{\gamma} \\
                          v'         & v          & J          \end{array}\right\}
\nonumber \\
&&\hspace*{-1cm}\times\bigg(
\left[\left[U_{\gamma}^{t''}(\mathbf{r})
\left[D_{n_{\gamma}L_{\gamma}}
  \rho_{v'}^{t'}(\bm{r},\bm{r}')\right]_J\right]_v\right]^t
-(-1)^{v'+v_{\gamma}-v}
(-1)^{t'+t''-t}
\left[\left[U_{\gamma}^{'t''}(\mathbf{r}')
\left[D'_{n_{\gamma}L_{\gamma}}
\rho_{v'}^{t'}(\bm{r},\bm{r}')\right]_J\right]_v\right]^t
\bigg)_{\bm{r}'=\bm{r}}=0.
\label{eq:13-1j15}
\end{eqnarray}
\end{widetext}
For our practical implementation of the CE
condition~(\ref{eq:13-1j15}), we proceed by transforming
the two differential operators, $D_{n_{\gamma}L_{\gamma}}$ and
$D'_{n_{\gamma}L_{\gamma}}$, which act on two different variables
$\bm{r}$ and $\bm{r}'$, respectively,
with the recoupling methods developed in Ref.~\cite{[Car10d]},
and we obtain,
\begin{eqnarray}
\label{eq:30a}
D_{n_{\gamma}L_{\gamma}}&=&\sum_{nLn'L'}K_{nLn'L'}^{n_{\gamma}L_{\gamma}}
(i)^{n}2^{-n'}\left[D_{n'L'}K_{nL}\right]_{L_{\gamma}}, \\
\label{eq:30b}
D'_{n_{\gamma}L_{\gamma}}&=&\sum_{nLn'L'}K_{nLn'L'}^{n_{\gamma}L_{\gamma}}
(-i)^{n}2^{-n'}\left[D_{n'L'}K_{nL}\right]_{L_{\gamma}}.
\end{eqnarray}
On the right-hand sides, operators $K_{nL}$ are the
higher-order spherical tensor derivatives~\cite{Raimondi1} built of
the relative momenta, $\bm{k}=(\bm{\nabla}-\bm{\nabla}')/2i$, and
operators $D_{n'L'}$ act on variable $\bm{r}$ {\em after} one sets
$\bm{r}'=\bm{r}$. The 91 numerical coefficients
$K_{nLn'L'}^{n_{\gamma}L_{\gamma}}$, which are needed up to
N$^{3}$LO, have been derived in Ref.~\cite{[Car10d]}. By using
Eqs.~(\ref{eq:30a}) and (\ref{eq:30b}), one can express the last line
of the Eq.~(\ref{eq:13-1j15}) as a linear combination of products of
pairs of secondary densities coupled in the spin and isospin spaces
to ranks $v$ and $t$, respectively. This final form, which for brevity is not
shown here explicitly, is used in obtaining the results of Sec.~\ref{sec:Results}.

\section{Continuity equations in the four spin-isospin channels}
\label{sec:Results}

Condition~(\ref{eq:13-1j15}) sets constraints on the coupling
constants $C_{a,\alpha}^{\beta,t}$ of the EDF. In our study,
these explicit constraints were obtained, with the aid of the
symbolic programming, as solutions of a linear system of equations,
where each equation is found by considering the coefficients standing at a
given product of pairs of secondary densities. Indices $(v,t)$
correspond to the choice of the channel under examination: (0,0) for
the CE in the scalar-isoscalar channel, (0,1) for the CE in the
scalar-isovector channel, (1,0) for the CE in the vector-isoscalar
channel, and (1,1) for the CE in the vector-isovector channel.

As a result of this analysis, for each spin-isospin channel we can
classify the coupling constants in four categories defined in
Ref.~\cite{Carlsson1}, namely, unrestricted, vanishing, independent, and
dependent. The unrestricted coupling constants are not affected by
condition~(\ref{eq:13-1j15}) and vanishing ones are forced by
this condition to be equal to zero. The remaining coupling constants
obey sets of linear conditions, whereby one can express the dependent
ones through independent ones. Obviously, for a given set of linear
conditions, this can be done in very many different ways; below we
present in each case only one choice thereof. We also use the name of
a free coupling constant to denote either the unrestricted or
independent one.

The structure of this section is as follows. First, in
Table~(\ref{generalTableCE}) we present for the four spin-isospin
channels an overview of results by showing the number of unrestricted
(U), vanishing (V), independent (I), and dependent (D) coupling
constants of the EDF. The sum of the unrestricted and independent
coupling constants gives the number of the free ones. Second, in
Sec.~\ref{zero-order-terms} we discuss the simplest case of the
zero-order terms, where one-body pseudopotentials reduce to simple
potential functions, and we can link our results to those presented
in the introductory Sec.~\ref{sec:the-CE1}. Next, in
Sec.~\ref{scalar_isoscalar} we briefly describe the results obtained
for the standard CE in the scalar-isoscalar channel. For the CEs in
the three other spin-isospin channel, we present our results in
Secs.~\ref{scalar_isovector}--\ref{vector_isovector}, where, for
clarity, only the second-order terms and general rules are discussed,
while the results for fourth and sixth orders are collected in
Appendices~\ref{app-scalar_isovector}--\ref{app-vector_isovector}.
\begin{table*}
\caption{\label{generalTableCE}Number of unrestricted (U),
vanishing (V), independent (I), and dependent (D) coupling constants
of different orders in the EDF up to N$^{3}$LO, shown for the four spin-isospin channels.
}
\begin{ruledtabular}
\begin{tabular}{cccccccccccccccccc}
&&\multicolumn{4}{c}{$v=0,~t=0$}&\multicolumn{4}{c}{$v=0,~t=1$}&\multicolumn{4}{c}{$v=1,~t=0$}&\multicolumn{4}{c}{$v=1,~t=1$}\\[-2ex]
&&\multicolumn{4}{c}{\hrulefill}&\multicolumn{4}{c}{\hrulefill}&\multicolumn{4}{c}{\hrulefill}&\multicolumn{4}{c}{\hrulefill}\\[-0.5ex]
    Order& Total& U&  V& I& D& U&  V& I& D& U&  V& I& D&U&  V&I&D  \\ \hline
        0&     4& 4&  0& 0& 0& 4&  0& 0& 0& 4&  0& 0& 0&1&  0&1&2  \\
        2&    24& 6&  0& 8&10& 3&  7& 6& 8& 2& 10& 4& 8&1& 11&1&11 \\
        4&    90& 6& 54& 6&24& 3& 57& 6&24& 2& 64& 4&20&1& 65&1&23 \\
        6&   258& 6&200& 6&46& 3&203& 6&46& 2&216& 4&36&1&217&1&39 \\
\hline
N$^{3}$LO&   376&22&254&20&80&13&267&18&78&10&290&12&64&4&293&4&75 \\
\end{tabular}
\end{ruledtabular}
\end{table*}

\subsection{\label{zero-order-terms}Constraints for zero-order terms}

For zero-order terms, condition~(\ref{eq:13-1j15}) gives no constraints
on the coupling constants, apart from those in the vector-isovector
channel, where we have,
\begin{subequations}
\begin{eqnarray}
C_{00,0000}^{0000,1}&=&\frac{1}{\sqrt{3}}C_{00,0011}^{0011,1}\label{eq:50zerothA}, \\
C_{00,0011}^{0011,0}&=&\frac{1}{\sqrt{3}}C_{00,0011}^{0011,1}\label{eq:50zerothB},
\end{eqnarray}
\end{subequations}
whereas the coupling constant $C_{00,0000}^{0000,0}$ is unrestricted.

It is interesting to discuss these results in connection with the
derivation of the CE for a spin-$\thalf$ particle moving in a local
potential, which we gave in Sec.~\ref{sec:the-CE1}. There, we
pointed out that the CE in the vector channel is valid when the
vector potential is parallel to the spin density. Exactly this
situation occurs for the zero-order EDF, where in each spin-isospin
channel the potential functions are simply proportional to densities.

In fact, the simple algebraic rule of the vector product in
Eq.~(\ref{eq:3j}) is equivalent to the coupling of pairs of
identical commuting rank 1 tensors to rank 1, which is identically
null. This is just the case for the coupling to rank $v=1$ ($t=1$) in
the spin (isospin) space for the vector-isoscalar (scalar-isovector)
channel of the CE, in which the identically null tensors formed by
the pairs of densities at zero order leave the corresponding coupling
constants unrestricted. However, when the coupling to a rank-1 tensor
is simultaneously performed in both spin and isospin space ($v=1$ and
$t=1$), as is the case in the vector-isovector channel of the CE, the
two negative signs in the commutation of the pair of densities give
an overall positive sign, and the above selection rule does not
apply. This explains why, at zero order of the vector-isovector
channel, condition~(\ref{eq:13-1j15}) does induce constraints on
coupling constants -- those given in Eqs.~(\ref{eq:50zerothA}) and
(\ref{eq:50zerothB}).

\subsection{\label{scalar_isoscalar}Constraints for the scalar-isoscalar channel}

For the standard CE in the scalar-isoscalar channel,
Eq.~(\ref{eq:21j}), the constraints that we found from
Eq.~(\ref{eq:13-1j15}) are exactly the same as those defining the
gauge-invariant functional up to N$^{3}$LO~\cite{Carlsson1}. This
constitutes an explicit verification of the general result presented
in Sec.~\ref{sec:the-CE2-00}. As in Ref.~\cite{Carlsson1}, we obtained
two sets of independent constraints, one for the isoscalar coupling
constants $C_{a,\alpha}^{\beta,0}$ and one for the isovector coupling
constants $C_{a,\alpha}^{\beta,1}$. Within each isospin channel, the
linear conditions for the scalar coupling constants, $C_{a,n_\alpha
L_\alpha 0 J_\alpha}^{n_\beta L_\beta 0 J_\beta,t}$, are disconnected
from the linear conditions for the vector coupling constants,
$C_{a,n_\alpha L_\alpha 1 J_\alpha}^{n_\beta L_\beta 1 J_\beta,t}$.

\subsection{Constraints for the scalar-isovector channel}
\label{scalar_isovector}

Validity of the CE for the scalar-isovector density, Eq.~(\ref{eq:702})
for $v=0$ and $t=1$,
imposes through Eq.~(\ref{eq:13-1j15}) certain specific constraints
on the coupling constants of the functional. At second order we obtain,
\begin{subequations}
\begin{eqnarray}
\label{eq:38a}
C_{00,2000}^{0000,t} & = & -\frac{3^t}{\sqrt{3}}C_{00,1101}^{1101,1-t}, \\
C_{00,1110}^{1110,t} & = & -\frac{3^t}{3\sqrt{3}}C_{00,2011}^{0011,1-t}-\frac{3^t}{3}\sqrt{\frac{5}{3}}C_{00,2211}^{0011,1-t}, \\
C_{00,1111}^{1111,t} & = & -\frac{3^t}{3}C_{00,2011}^{0011,1-t}+\frac{3^t}{6}\sqrt{5}C_{00,2211}^{0011,1-t}, \\
\label{eq:38b}
C_{00,1112}^{1112,t} & = & -\frac{3^t}{3}\sqrt{\frac{5}{3}}C_{00,2011}^{0011,1-t}-\frac{3^t}{6\sqrt{3}}C_{00,2211}^{0011,1-t}, \\
\label{eq:38c}
C_{11,1111}^{0000,t} & = & C_{11,0011}^{1101,t} = 0 , \\
\label{eq:38d}
C_{20,0000}^{0000,1} & = &
C_{20,0011}^{0011,1} =
C_{22,0011}^{0011,1} = 0 .
\end{eqnarray}
\end{subequations}

Constraints in Eqs.~(\ref{eq:38a})--(\ref{eq:38b}) connect the
isoscalar and isovector coupling constants. The numerical
coefficients of the corresponding linear combinations are the same as
those for the scalar-isoscalar CE, see Eqs.~(C1)--(C4) of
Ref.~\cite{Carlsson1}, apart from factors of $\sqrt{3}$ explained
before Eq.~(\ref{eq:hfeq2}). However, conditions for the
scalar-isoscalar CE keep the coupling constants in the two isospin channels disconnected. Moreover,
in the scalar-isovector channel the spin-orbit coupling constants
must vanish, Eq.~(\ref{eq:38c}), along with the isovector surface
coupling constants, Eq.~(\ref{eq:38d}). On the other hand, the
corresponding isoscalar surface coupling constants,
$C_{20,0000}^{0000,0}$, $C_{20,0011}^{0011,0}$, and
$C_{22,0011}^{0011,0}$, are left unrestricted.

For the fourth and sixth orders, analogous constraints are
presented in Appendix~\ref{app-scalar_isovector}.

\subsection{Constraints for the vector-isoscalar channel}
\label{vector_isoscalar}

Validity of the CE for the vector-isoscalar density,
Eq.~(\ref{eq:702}) for $v=1$ and $t=0$, imposes through
Eq.~(\ref{eq:13-1j15}) at second order the following constraints on
the coupling constants of the functional,
\begin{subequations}
\begin{eqnarray}
C_{00,1101}^{1101,t} & = & -\frac{1}{\sqrt{3}}C_{00,2011}^{0011,t},\label{eq:44a}\\
C_{00,1110}^{1110,t} & = & -\frac{1}{\sqrt{3}}C_{00,2000}^{0000,t}, \\
C_{00,1111}^{1111,t} & = & -C_{00,2000}^{0000,t}, \\
C_{00,1112}^{1112,t} & = & -\sqrt{\frac{5}{3}}C_{00,2000}^{0000,t},\\
C_{20,0011}^{0011,t} & = & C_{22,0011}^{0011,t} = 0,
\label{eq:44e} \\
C_{11,1111}^{0000,t} & = & C_{11,0011}^{1101,t} = 0,
\label{eq:44d} \\
C_{00,2211}^{0011,t} & = & 0 ,\label{eq:44b}
\end{eqnarray}
\end{subequations}
whereas the two coupling constants $C_{20,0000}^{0000,t}$ are left
unrestricted. We note here that the constraints now connect scalar
and vector coupling constants. Altogether, at second order, for the
vector-isoscalar channel of the CE we have 6 free and 8 dependent
coupling constants. Apart from that, 10
second-order coupling constants must vanish, which includes the
surface ones in Eq.~(\ref{eq:44e}), spin-orbit ones of the
Eq.~(\ref{eq:44d}), and tensor ones in Eq.~(\ref{eq:44b}).

For the fourth and sixth orders, analogous constraints are
presented in Appendix~\ref{app-vector_isoscalar}.

\subsection{Constraints for the vector-isovector channel}
\label{vector_isovector}

Validity of the CE for the vector-isovector density,
Eq.~(\ref{eq:702}) for $v=1$ and $t=1$, imposes through
Eq.~(\ref{eq:13-1j15}) constraints on the coupling constants that
relate them in both spin and isospin spaces. At all the orders, we
can express all dependent coupling constants through only one vector
coupling constant, which can be chosen either in the set of the
vector-isoscalar or vector-isovector ones. At second order, these
constraints read,
\begin{subequations}
\begin{eqnarray}
C_{00,2000}^{0000,t} & = & -\sqrt{3}(\sqrt{3})^{t}C_{00,1110}^{1110,0},\\
C_{00,1101}^{1101,t} & = & \sqrt{3}(\sqrt{3})^{t}C_{00,1110}^{1110,0},\\
C_{00,1110}^{1110,1} & = & \sqrt{3}C_{00,1110}^{1110,0}, \\
C_{00,1111}^{1111,t} & = & \sqrt{3}(\sqrt{3})^{t}C_{00,1110}^{1110,0}, \\
C_{00,1112}^{1112,t} & = & \sqrt{5}(\sqrt{3})^{t}C_{00,1110}^{1110,0}, \\
C_{00,2011}^{0011,t} & = & -3(\sqrt{3})^{t}C_{00,1110}^{1110,0}, \\
C_{20,0011}^{0011,t} & = & C_{22,0011}^{0011,t} = 0,
\label{eq:50a00} \\
C_{20,0000}^{0000,1} & = &  0,
\label{eq:50a01} \\
C_{00,2211}^{0011,t} & = & 0,\label{eq:50a}\\
C_{11,1111}^{0000,t} & = & C_{11,0011}^{1101,t} = 0
\label{eq:50b},
\end{eqnarray}
\end{subequations}
and only coupling constant $C_{20,0000}^{0000,0}$ is unrestricted.
Altogether, at second order for the vector-isovector channel of the
CE we have 2 free coupling constants and 11 coupling constants that
are dependent. All the remaining 11 second-order coupling constants,
which includes the
surface ones in Eqs.~(\ref{eq:50a00})--(\ref{eq:50a01}), the tensor ones in Eq.~(\ref{eq:50a}) and spin-orbit
ones in Eq.~(\ref{eq:50b}), are forced to be equal to zero.

For the fourth and sixth orders, analogous constraints are
presented in Appendix~\ref{app-vector_isovector}.

\section{\label{sec:Conclusions}conclusions}

In the present work, we have derived sets of constraints on the
coupling constants of the N$^{3}$LO energy density functional that
guarantee the validity of the continuity equation in the four
spin-isospin channels. In the scalar-isoscalar channel, these
constraints are identical to those induced by the standard local
gauge invariance conditions. We extended this connection to vector
and isovector channels, where the validity of the continuity
equations is equivalent to the local gauge invariance with respect to
spin and isospin rotations, respectively.

We note here that in our analysis we implicitly assumed that all
densities that build the N$^{3}$LO energy density functional are
nonzero. Of course, there can be many specific situations when some
densities vanish, and thus the coupling constants related to them
become unrestricted. This occurs, for instance, when densities are
restricted by some symmetry conditions. Obviously, the methods
developed in the present work can then be applied to derive new
(weaker) sets of constraints that correspond to each one particular
case. For example, when the proton-neutron symmetry is conserved,
see, e.g., Refs.~\cite{[Per04],[Roh10]}, the $\pm1$ components of all
isovector densities ($v=1$) vanish, and the energy density is
invariant with respect to a one-dimensional U(1) gauge rotation in
the isospin space, that is, with respect to
$U=\exp(i\gamma_{10}\tau^{10})$. Then, the continuity equations for
neutrons and protons decouple from one another and become
independently valid, provided the isoscalar and isovector coupling
constants independently obey the standard gauge-invariance
conditions.

\acknowledgments

This work was supported in part by the Academy of Finland and the
University of Jyv\"askyl\"a within the FIDIPRO programme, and by the
Polish Ministry of Science and Higher Education.

\appendix

\section{Constraints for the scalar-isovector channel (fourth and sixth orders)}
\label{app-scalar_isovector}

At fourth order we found that the isoscalar and isovector coupling
constants are connected or not depending on the parity of quantum
numbers $n$. Moreover, similarly as for the scalar-isoscalar channel of the CE,
the scalar and vector coupling constants are kept apart. The
constraints among the scalar coupling constants read,
\begin{eqnarray}
\label{app-eq:40}
C_{00,4000}^{0000,t}&=& \frac{3 C_{00,2202}^{2202,t}}{2 \sqrt{5}},\\
C_{00,2000}^{2000,t}&=& \frac{1}{2} \sqrt{5} C_{00,2202}^{2202,t},\\
C_{00,3101}^{1101,t}&=& -2\sqrt{\frac{3}{5}}3^t C_{00,2202}^{2202,1-t},
\end{eqnarray}
and those among the vector coupling constants read,
\begin{eqnarray}
C_{00,3110}^{1110,t}&=& -\frac{2}{\sqrt{5}} 3^t C_{00,2212}^{2212,1-t}-\frac{7}{\sqrt{15}} 3^t C_{00,4211}^{0011,1-t},\\
C_{00,3111}^{1111,t}&=& -2  (3^t)\sqrt{\frac{3}{5}} C_{00,2212}^{2212,1-t},\\
C_{00,3112}^{1112,t}&=& -2 (3^t)C_{00,2212}^{2212,1-t}-\frac{14}{5}\frac{3^t}{\sqrt{3}} C_{00,4211}^{0011,1-t},\\
C_{00,3312}^{1112,t}&=&-\frac{2}{3^{1-t}} \sqrt{\frac{7}{5}} C_{00,4211}^{0011,1-t},\\
C_{00,4011}^{0011,t}&=& \frac{3}{2} \sqrt{\frac{3}{5}} C_{00,2212}^{2212,t}+\frac{7 C_{00,4211}^{0011,t}}{4 \sqrt{5}},\\
C_{00,2011}^{2011,t}&=& \frac{1}{2} \sqrt{15} C_{00,2212}^{2212,t}+\frac{7}{12} \sqrt{5} C_{00,4211}^{0011,t},\\
C_{00,2211}^{2011,t}&=& \frac{7}{3} C_{00,4211}^{0011,t},\\
C_{00,2211}^{2211,t}&=& \sqrt{\frac{3}{5}} C_{00,2212}^{2212,t}+\frac{7 C_{00,4211}^{0011,t}}{3 \sqrt{5}},\\
C_{00,2213}^{2213,t}&=& \sqrt{\frac{7}{5}} C_{00,2212}^{2212,t}+\frac{1}{2} \sqrt{\frac{21}{5}} C_{00,4211}^{0011,t}
\label{app-eq:40bis}.
\end{eqnarray}

We also found that the fourth-order surface isovector coupling constants
must vanish,
\begin{eqnarray}
C_{40,0000}^{0000,1} &=&
C_{40,0011}^{0011,1} =
C_{42,0011}^{0011,1} = 0,
\label{app-eq:41c}
\end{eqnarray}
whereas the corresponding isoscalar coupling constants are
unrestricted. Apart from the coupling constants discussed above, all
the remaining 54 fourth-order coupling constants are forced to be
equal to zero.

In the same way, at sixth order we found the following constraints for
the scalar,
\begin{eqnarray}
C_{00,6000}^{0000,t}&=& -3^t\frac{3 C_{00,3303}^{3303,1-t}}{4 \sqrt{7}},\\
C_{00,4000}^{2000,t}&=& -3^t\frac{3}{4} \sqrt{7} C_{00,3303}^{3303,1-t},\\
C_{00,4202}^{2202,t}&=& -3^t3 \sqrt{\frac{5}{7}} C_{00,3303}^{3303,1-t},\\
C_{00,5101}^{1101,t}&=& \frac{9}{2} \sqrt{\frac{3}{7}} C_{00,3303}^{3303,t},\\
C_{00,3101}^{3101,t}&=& \frac{9}{10} \sqrt{21} C_{00,3303}^{3303,t},
\label{app-eq:42}
\end{eqnarray}
and vector coupling constants,
\begin{eqnarray}
C_{00,5110}^{1110,t}&=& -3^t\frac{C_{00,4212}^{2212,1-t}}{2 \sqrt{5}}-3^t3 \sqrt{\frac{3}{5}} C_{00,6211}^{0011,1-t}\label{app-eq:42bisA},\\
C_{00,5111}^{1111,t}&=& -3^t\frac{1}{2} \sqrt{\frac{3}{5}} C_{00,4212}^{2212,1-t},\\
C_{00,5112}^{1112,t}&=& -3^t\frac{1}{2} C_{00,4212}^{2212,1-t}-3^t\frac{6}{5} \sqrt{3} C_{00,6211}^{0011,1-t},\\
C_{00,5312}^{1112,t}&=& -\frac{4}{3^{1-t}} \sqrt{\frac{7}{5}} C_{00,6211}^{0011,1-t},\\
C_{00,3110}^{3110,t}&=& -3^t\frac{7 C_{00,4212}^{2212,1-t}}{10 \sqrt{5}}-3^t\frac{21}{5} \sqrt{\frac{3}{5}} C_{00,6211}^{0011,1-t},\\
C_{00,3111}^{3111,t}&=& -3^t\frac{7}{10} \sqrt{\frac{3}{5}} C_{00,4212}^{2212,1-t},\\
C_{00,3112}^{3112,t}&=& -3^t\frac{7}{10} C_{00,4212}^{2212,1-t}-3^t\frac{42}{25} \sqrt{3} C_{00,6211}^{0011,1-t},\\
C_{00,3312}^{3112,t}&=& -3^t\frac{12}{5} \sqrt{\frac{7}{5}} C_{00,6211}^{0011,1-t},\\
C_{00,3312}^{3312,t}&=& -3^t\frac{1}{9} C_{00,4212}^{2212,1-t}-3^t\frac{2}{5} \sqrt{3} C_{00,6211}^{0011,1-t},\\
C_{00,3313}^{3313,t}&=& -\frac{3^t}{9} \sqrt{\frac{7}{5}} C_{00,4212}^{2212,1-t},\\
C_{00,3314}^{3314,t}&=& -\frac{C_{00,4212}^{2212,1-t}}{3^{1-t} \sqrt{5}}-\frac{8 C_{00,6211}^{0011,1-t}}{3^{1-t} \sqrt{15}},\\
C_{00,6011}^{0011,t}&=& \frac{1}{4} \sqrt{\frac{3}{5}} C_{00,4212}^{2212,t}+\frac{3 C_{00,6211}^{0011,t}}{2 \sqrt{5}}\label{app-eq:4200A},\\
C_{00,4011}^{2011,t}&=& \frac{7}{4} \sqrt{\frac{3}{5}} C_{00,4212}^{2212,t}+\frac{21 C_{00,6211}^{0011,t}}{2 \sqrt{5}},\\
C_{00,4211}^{2011,t}&=& 6 C_{00,6211}^{0011,t},\\
C_{00,4011}^{2211,t}&=& \frac{21}{5} C_{00,6211}^{0011,t},\\
C_{00,4211}^{2211,t}&=& \sqrt{\frac{3}{5}} C_{00,4212}^{2212,t}+\frac{12 C_{00,6211}^{0011,t}}{\sqrt{5}},\\
C_{00,4213}^{2213,t}&=& \sqrt{\frac{7}{5}} C_{00,4212}^{2212,t}+18 \sqrt{\frac{3}{35}} C_{00,6211}^{0011,t},\\
C_{00,4413}^{2213,t}&=& \frac{4 C_{00,6211}^{0011,t}}{\sqrt{5}}
\label{app-eq:42bisB}.
\end{eqnarray}

We also found that the sixth-order surface isovector coupling constants
must vanish,
\begin{eqnarray}
C_{60,0000}^{0000,1} &=&
C_{60,0011}^{0011,1} =
C_{62,0011}^{0011,1} = 0,
\label{app-eq:43c}
\end{eqnarray}
whereas the corresponding isoscalar coupling constants are
unrestricted. Apart from the coupling constants discussed above, all
the remaining 200 sixth-order coupling constants are forced to be
equal to zero.

We have seen that for the scalar-isovector channel, the coupling
constants are diagonal or nondiagonal in the isospin quantum number
$t$. We can understand this point considering the fact that in order
to separate the scalar-isovector channel of the CE and obtain
condition~(\ref{eq:13-1j15}) for $t=1$, we have to multiply
Eq.~(\ref{eq:7c}) by the isospin operator $\tau_{\tau'\tau}$. Then, the
isospin index $t''$ tells us in which half of the isospin
space the coupling constants is. Nondiagonal constraints mean, in
fact, that the same pair of secondary densities in the final form of
condition~(\ref{eq:13-1j15}) can be produced by two terms of the
functional that are isoscalar and isovector. This
is possible, because the coupling to rank $t=1$ allows for pairs
of densities nondiagonal in the isospin space, and this, in turn, boils
down to constraints for coupling constants nondiagonal in the isospin space.

\section{Constraints for the vector-isoscalar channel (fourth and sixth orders)}
\label{app-vector_isoscalar}

At fourth order, we again found two identical sets of linear
combinations of the isoscalar and isovector coupling constants, in which
the scalar and vector coupling constants are in same case connected
to one another, that is,
\begin{eqnarray}
C_{00,4000}^{0000,t}&=& \frac{3 C_{00,2202}^{2202,t}}{2 \sqrt{5}},\label{app-eq:45a}\\
C_{00,2000}^{2000,t}&=& \frac{1}{2} \sqrt{5} C_{00,2202}^{2202,t},\\
C_{00,3101}^{1101,t}&=& -\frac{6 C_{00,2212}^{2212,t}}{\sqrt{5}},\\
C_{00,3110}^{1110,t}&=& -2 \sqrt{\frac{3}{5}} C_{00,2202}^{2202,t},\\
C_{00,3111}^{1111,t}&=& -\frac{6 C_{00,2202}^{2202,t}}{\sqrt{5}},\\
C_{00,3112}^{1112,t}&=& -2 \sqrt{3} C_{00,2202}^{2202,t},\\
C_{00,4011}^{0011,t}&=& \frac{3}{2} \sqrt{\frac{3}{5}} C_{00,2212}^{2212,t},\\
C_{00,2011}^{2011,t}&=& \frac{1}{2} \sqrt{15} C_{00,2212}^{2212,t},\\
C_{00,2211}^{2211,t}&=& \sqrt{\frac{3}{5}} C_{00,2212}^{2212,t},\\
C_{00,2213}^{2213,t}&=& \sqrt{\frac{7}{5}} C_{00,2212}^{2212,t},\\
C_{00,2211}^{2011,t}&=& C_{00,3312}^{1112,t}=C_{00,4211}^{0011,t}=0,
\label{app-eq:45l}
 \end{eqnarray}
with the two coupling constants $C_{40,0000}^{0000,t}$ left unrestricted.
Apart from these 6 free and 20 dependent coupling constants, the
vector-isoscalar channel of the CE requires that all the remaining
fourth-order coupling constants are forced to be equal to zero. In particular
in the Eq.~(\ref{app-eq:45l}) we showed the vanishing coupling constants,
belonging to the set of ones with indices $m=0$ and $I=0$, which were found
to be non-vanishing in the scalar-isoscalar channel.

At sixth order the pattern of the results is the same, and we have,
\begin{eqnarray}
C_{00,3303}^{3303,t}&=& -\frac{1}{3} \sqrt{\frac{7}{15}} C_{00,4212}^{2212,t},\\
C_{00,6000}^{0000,t}&=& -\frac{C_{00,5112}^{1112,t}}{2 \sqrt{15}},\\
C_{00,4000}^{2000,t}&=& -\frac{7 C_{00,5112}^{1112,t}}{2 \sqrt{15}},\\
C_{00,4202}^{2202,t}&=& -\frac{2 C_{00,5112}^{1112,t}}{\sqrt{3}},\\
C_{00,5101}^{1101,t}&=& -\frac{3 C_{00,4212}^{2212,t}}{2 \sqrt{5}},\\
C_{00,3101}^{3101,t}&=& -\frac{21 C_{00,4212}^{2212,t}}{10 \sqrt{5}},\\
C_{00,5110}^{1110,t}&=& \frac{C_{00,5112}^{1112,t}}{\sqrt{5}},\\
C_{00,5111}^{1111,t}&=& \sqrt{\frac{3}{5}} C_{00,5112}^{1112,t},\\
C_{00,3110}^{3110,t}&=& \frac{7 C_{00,5112}^{1112,t}}{5 \sqrt{5}},\\
C_{00,3111}^{3111,t}&=& \frac{7}{5} \sqrt{\frac{3}{5}} C_{00,5112}^{1112,t},\\
C_{00,3112}^{3112,t}&=& \frac{7}{5} C_{00,5112}^{1112,t},\\
C_{00,3312}^{3312,t}&=& \frac{2}{9} C_{00,5112}^{1112,t},\\
C_{00,3313}^{3313,t}&=& \frac{2}{9} \sqrt{\frac{7}{5}} C_{00,5112}^{1112,t},\\
C_{00,3314}^{3314,t}&=& \frac{2 C_{00,5112}^{1112,t}}{3 \sqrt{5}},\\
C_{00,6011}^{0011,t}&=& \frac{1}{4} \sqrt{\frac{3}{5}} C_{00,4212}^{2212,t},\\
C_{00,4011}^{2011,t}&=& \frac{7}{4} \sqrt{\frac{3}{5}} C_{00,4212}^{2212,t},\\
C_{00,4211}^{2211,t}&=& \sqrt{\frac{3}{5}} C_{00,4212}^{2212,t},\\
C_{00,4213}^{2213,t}&=& \sqrt{\frac{7}{5}} C_{00,4212}^{2212,t},\\
C_{00,4011}^{2211,t}&=&C_{00,4413}^{2213,t}=C_{00,4211}^{2011,t}= 0,\label{app-eq:46a}\\
C_{00,3312}^{3112,t}&=& C_{00,5312}^{1112,t}=C_{00,6211}^{0011,t}=0
 \label{app-eq:46b},
\end{eqnarray}
with the two coupling constants $C_{60,0000}^{0000,t}$ left unrestricted.
Apart from these 6 free and 36 dependent coupling constants, the
vector-isoscalar channel of the CE requires that all the remaining
sixth-order coupling constants are forced to be equal to zero. As before, we
listed explicitely the vanishing coupling constants (see Eqs.~(\ref{app-eq:46a})--(\ref{app-eq:46b})),
 which were found
to be non-vanishing in the scalar-isoscalar channel.

The general rule that we have specified at the end of the
Appendix~\ref{app-scalar_isovector}, can be applied now to explain
the results of this section, where at all the orders we found
constraints that are nondiagonal in the spin space. Here, the reason
is the possibility of having pairs of secondary densities that are
coupled to rank $v=1$. These pairs can appear at both scalar and
vector coupling constants, which results in relating them to one
another.

\section{Constraints for the vector-isovector channel (fourth and sixth orders)}
\label{app-vector_isovector}

At fourth order we found the following constraints,
\begin{eqnarray}
C_{00,4000}^{0000,t}&=& (\sqrt{3})^{t}\frac{C_{00,4011}^{0011,0}}{\sqrt{3}},\\
C_{00,2000}^{2000,t}&=& (\sqrt{3})^{t}\frac{5 C_{00,4011}^{0011,0}}{3 \sqrt{3}},\\
C_{00,3101}^{1101,t}&=& -(\sqrt{3})^{t}\frac{4 C_{00,4011}^{0011,0}}{\sqrt{3}},\\
C_{00,2202}^{2202,t}&=& (\sqrt{3})^{t}\frac{2}{3} \sqrt{\frac{5}{3}} C_{00,4011}^{0011,0},\\
C_{00,3110}^{1110,t}&=& -(\sqrt{3})^{t}\frac{4}{3} C_{00,4011}^{0011,0},\\
C_{00,3111}^{1111,t}&=& -(\sqrt{3})^{t}\frac{4 C_{00,4011}^{0011,0}}{\sqrt{3}},\\
C_{00,3112}^{1112,t}&=& -(\sqrt{3})^{t}\frac{4}{3} \sqrt{5} C_{00,4011}^{0011,0},\\
C_{00,4011}^{0011,1}&=& \sqrt{3} C_{00,4011}^{0011,0},\\
C_{00,2011}^{2011,t}&=& (\sqrt{3})^{t}\frac{5}{3} C_{00,4011}^{0011,0},\\
C_{00,2211}^{2211,t}&=& (\sqrt{3})^{t}\frac{2}{3} C_{00,4011}^{0011,0},\\
C_{00,2212}^{2212,t}&=& (\sqrt{3})^{t}\frac{2}{3} \sqrt{\frac{5}{3}} C_{00,4011}^{0011,0},\\
C_{00,2213}^{2213,t}&=& (\sqrt{3})^{t}\frac{2}{3} \sqrt{\frac{7}{3}} C_{00,4011}^{0011,0},\\
C_{00,2211}^{2011,t}&=&C_{00,3312}^{1112,t}=C_{00,4211}^{0011,t}= 0
\label{app-eq:51},
 \end{eqnarray}
and only coupling constant $C_{40,0000}^{0000,0}$ is unrestricted.
Apart from these 2 free and 23 dependent coupling constants, the
vector-isovector channel of the CE requires that all the remaining
fourth-order coupling constants are forced to be equal to zero. In particular
in the Eq.~(\ref{app-eq:51}) we showed the vanishing coupling constants, which were found
to be non-vanishing in the scalar-isoscalar channel.

At sixth order we have,
\begin{eqnarray}
C_{00,6000}^{0000,t}&=& -(\sqrt{3})^{t}\frac{C_{00,5112}^{1112,0}}{2 \sqrt{15}},\\
C_{00,4000}^{2000,t}&=& -(\sqrt{3})^{t}\frac{7 C_{00,5112}^{1112,0}}{2 \sqrt{15}},\\
C_{00,4202}^{2202,t}&=& -(\sqrt{3})^{t}\frac{2 C_{00,5112}^{1112,0}}{\sqrt{3}},\\
C_{00,5101}^{1101,t}&=& (\sqrt{3})^{t}\sqrt{\frac{3}{5}} C_{00,5112}^{1112,0},\\
C_{00,3101}^{3101,t}&=& (\sqrt{3})^{t}\frac{7}{5} \sqrt{\frac{3}{5}} C_{00,5112}^{1112,0},\\
C_{00,3303}^{3303,t}&=& (\sqrt{3})^{t}\frac{2}{9} \sqrt{\frac{7}{5}} C_{00,5112}^{1112,0},\\
C_{00,5110}^{1110,t}&=& (\sqrt{3})^{t}\frac{C_{00,5112}^{1112,0}}{\sqrt{5}},\\
C_{00,5111}^{1111,t}&=& (\sqrt{3})^{t}\sqrt{\frac{3}{5}} C_{00,5112}^{1112,0},\\
C_{00,5112}^{1112,1}&=& \sqrt{3} C_{00,5112}^{1112,0},\\
C_{00,3110}^{3110,t}&=& (\sqrt{3})^{t}\frac{7 C_{00,5112}^{1112,0}}{5 \sqrt{5}},\\
C_{00,3111}^{3111,t}&=& (\sqrt{3})^{t}\frac{7}{5} \sqrt{\frac{3}{5}} C_{00,5112}^{1112,0},\\
C_{00,3112}^{3112,t}&=& (\sqrt{3})^{t}\frac{7}{5} C_{00,5112}^{1112,0},\\
C_{00,3312}^{3312,t}&=& (\sqrt{3})^{t}\frac{2}{9} C_{00,5112}^{1112,0},\\
C_{00,3313}^{3313,t}&=& (\sqrt{3})^{t}\frac{2}{9} \sqrt{\frac{7}{5}} C_{00,5112}^{1112,0},\\
C_{00,3314}^{3314,t}&=& (\sqrt{3})^{t}\frac{2 C_{00,5112}^{1112,0}}{3 \sqrt{5}},\\
C_{00,6011}^{0011,t}&=& -(\sqrt{3})^{t}\frac{C_{00,5112}^{1112,0}}{2 \sqrt{5}},\\
C_{00,4011}^{2011,t}&=& -(\sqrt{3})^{t}\frac{7 C_{00,5112}^{1112,0}}{2 \sqrt{5}},\\
C_{00,4211}^{2211,t}&=& -(\sqrt{3})^{t}\frac{2 C_{00,5112}^{1112,0}}{\sqrt{5}},\\
C_{00,4213}^{2213,t}&=& -(\sqrt{3})^{t}2 \sqrt{\frac{7}{15}} C_{00,5112}^{1112,0},\\
C_{00,4212}^{2212,t}&=& -(\sqrt{3})^{t}\frac{2 C_{00,5112}^{1112,0}}{\sqrt{3}},\\
C_{00,4011}^{2211,t}&=&C_{00,4413}^{2213,t}=C_{00,3312}^{3112,t}= 0,\label{app-eq:52a}\\
C_{00,4211}^{2011,t}&=&C_{00,6211}^{0011,t}=C_{00,5312}^{1112,t}=0
\label{app-eq:52b},
 \end{eqnarray}
and only coupling constant $C_{60,0000}^{0000,0}$ is unrestricted.
Apart from 2 free and 39 dependent coupling constants, the
vector-isovector channel of the CE requires that all the remaining
sixth-order coupling constants are forced to be equal to zero. In particular,
in the Eqs.~(\ref{app-eq:52a})--(\ref{app-eq:52b}) we showed the vanishing coupling constants that were found
to be non-vanishing in the scalar-isoscalar channel.

The results presented in this section show simultaneously both features we saw
respectively in Appendices~\ref{app-scalar_isovector} and
\ref{app-vector_isoscalar}. At all
orders, one can express all coupling constants through only one independent
coupling constant, in such a way that the constraints are
nondiagonal in both spin and isospin space. Again, this fact is due
to the rank of $v=1$ and $t=1$ in the pairs of densities in the final
form of condition~(\ref{eq:13-1j15}), which allows the coupling
constants at different spins and isospins to enter into the same
constraints.


\end{document}